\documentclass[prb,twocolumn]{revtex4} %CM

\usepackage{graphicx}
\usepackage{color}
\usepackage{subfigure}
\usepackage{verbatim}
\def\Gd3+{Gd$^{3+}$}
\def\LaOFFeAs{LaO$_{1-x}$F$_{x}$FeAs}

\def\GdOFFeAs{GdO$_{1-x}$F$_{x}$FeAs}

\def\Tsdw{\emph{T$_{SDW}$}}
\def\Tst{\emph{T$_{ST}$}}
\def\Tc{\emph{T$_{c}$}}
\def\Tif{\emph{T$_{if}$}}

\def\Hrespp{$H_{res\perp}$}
\def\Hrespl{$H_{res\parallel}$}
\def\Hext{$H_{ext}$}

\def\Hintpp{$H_{int \perp}$}
\def\Hintpl{$H_{int \parallel}$}

\def\cp {$c_{\rm p}$}
\def\mb {$\mu_{\rm B}$}
\newcommand{\figref}[1]{Fig.\,\protect\ref{#1}}

\newcommand{\gdo}{GdOFeAs}
\newcommand{\lao}{LaOFeAs}

\newcommand{\tngd}{$T_{\rm{N}^{Gd}}$}

\begin{document}
\title{High-field electron spin resonance spectroscopy study of GdO$_{1-x}$F$_{x}$FeAs superconductors}
%\author{A. Author,\textsuperscript{1,*} S. Author,\textsuperscript{2} and T. Author\textsuperscript{1}}
\author{A. Alfonsov$^1$, F. Mur\'{a}nyi$^2$, V. Kataev$^1$, G. Lang$^1$, N. Leps$^1$, L.
Wang$^1$, R. Klingeler$^3$, A. Kondrat$^1$, C. Hess$^1$, S.
Wurmehl$^1$, A. K\"ohler$^1$, G. Behr$^1$, S. Hampel$^1$, M.
Deutschmann$^1$, S. Katrych$^4$, N. D. Zhigadlo$^4$, Z.
Bukowski$^4$, J. Karpinski$^4$, and B. B\"{u}chner$^1$}
\affiliation{$^1$ IFW Dresden, Institute for Solid State Research,
01069 Dresden, Germany \\
$^2$ Universit\"{a}t Z\"{u}rich - Physik-Institut, CH-8057
Z\"{u}rich, Switzerland\\
$^3$ Kirchhoff Institute for Physics, Heidelberg University, D-69120 Heidelberg, Germany\\
$^4$ Laboratory for Solid State Physics, ETH Zurich, CH-8093 Zurich,
Switzerland }

\begin{abstract}

We report a detailed investigation of \GdOFFeAs\ ($x=0$, $0.07$ and
$0.14$) samples by means of high-field/high-frequency electron spin
resonance (HF-ESR) together with measurements of thermodynamic and
transport properties. The parent GdOFeAs compound exhibits Fe
long-range magnetic order below 128\,K, whereas both doped samples
do not show such order and are superconducting with $\Tc=20$\,K
($x=0.07$) and $\Tc=45$\,K ($x=0.14$). The \Gd3+\ HF-ESR reveals an
appreciable exchange coupling between Gd and Fe moments, through
which the static magnetic order is clearly seen in the parent
compound. Owing to this coupling, HF-ESR can probe sensitively the
evolution of the magnetism in the FeAs planes upon F doping. It is
found that in both superconducting samples, where the Fe long-range
order is absent, there are short-range, static on the ESR time scale
magnetic correlations between Fe spins. Their occurrence on a large
doping scale may be indicative of the ground states' coexistence.

\end{abstract}

\maketitle

\section{Introduction}

Iron-pnictide superconductors\cite{kamihara_feas} with
superconducting critical temperatures up to
55\,K\cite{Chen2008_feas, Drew2009, Hess2009} have attracted a huge
interest due to striking similarities to superconducting cuprates as
well as due to their original properties. Indeed, most families of
these layered materials feature an antiferromagnetically (AFM)
ordered parent compound, and the evolution of superconductivity
concomitantly with suppression of AFM order upon doping. However
there are important differences which render the Fe-pnictides a
separate new class of superconducting materials. Most striking of
them are semi-metallicity and the spin density wave (SDW) character
of the AFM order in the undoped pnictides contrasted with the
Mott-insulating AFM state in the cuprates, as well as a multi-band
versus single-band electronic structure in the Fe-pnictide and
cuprate high-temperature superconductors, respectively.

Beyond study of the superconducting ground state, and of the
magnetic and associated structural transitions seen in the parent
compound, much attention has been devoted to the issue of the ground
states' coexistence. Discrepancies on this issue have been found
between different families\cite{Drew2009, Luetkens2009, Zhao2008,
Sanna2010, Julien2009, Laplace2009, Urbano2010, Carlo2009}, with the
variation of the boundary of the two ground states and different
length scales of coexistence, especially in the so-called 1111
family. In this family, which has the composition $R$OFe$Pn$ ($R$ -
rare earth, $Pn$ - pnictide), the superconductivity evolves with the
substitution of fluorine for oxygen. Here, replacement of one rare
earth element with another can cause a significant variation of
properties. Whereas in La-based \emph{superconducting} samples there
is evidence against static magnetic order in the FeAs
planes\cite{Luetkens2009}, in the case of superconducting samples
based on different magnetic rare earths ($R=$ Sm, Nd, Ce) evidence
of remanent static magnetism is found \cite{Drew2009, Carlo2009}.
The situation appears complicated due to the fact that the magnetism
then tends to be of a short-range order or disordered, possibly even
dynamic\cite{Bernhard2009}, which calls for the use of local probe
techniques. These two different pictures complicate the
establishment of the unified phase diagram for 1111 pnictides,
necessary for the full understanding of these materials. In
addition, as was shown by
NMR\cite{nakai_FeAs_LaNMR,PhysRevB.81.100508} and
$\mu$SR\cite{maeter_feas_musr_2009} studies, there is a magnetic
coupling between 4$f$ (Ce, Pr and Sm) and 3$d$ (Fe) moments. Such
coupling of the rare earth to the FeAs plane might give an
additional contribution to the difference in physical properties of
different 1111-type superconductors.

In the present work we investigate the evolution of the magnetism
upon fluorine doping in Gd-based 1111 compound by means of high
field/high frequency electron spin resonance (HF-ESR) complemented
with measurements of thermodynamic and transport properties. The ESR
data reveal a significant exchange coupling of Gd- and Fe-moments in
the parent GdOFeAs sample which enables the \Gd3+\ HF-ESR to probe
sensitively the formation of the static SDW magnetic order in the
FeAs planes. Interestingly, it is found that the signatures of such
an order are still observed in the ESR spectra after doping. In
particular, though long-range SDW order present at very low doping
is suppressed at doping levels where superconductivity appears, our
results imply static on the ESR time scale, likely short-range,
magnetic correlations between Fe spins. This result suggests that
\GdOFFeAs\ compounds may feature coexistence of quasi-static
magnetism and superconductivity on a large doping range.

\section{Experimental}

\subsection{Setups}

The magnetization has been studied by means of a commercial SQUID
magnetometer (MPMS-XL5, Quantum Design). For the thermal expansion
measurements a capacitance dilatometer was utilized, which allows a
very accurate study of crystal length changes. We measured the
macroscopic length changes $dL/L$ of polycrystalline samples. The
linear thermal expansion coefficient $\alpha$ was calculated as the
first temperature derivative of $dL/L$, while the volume thermal
expansion coefficient is given by $\beta_\mathrm{vol}=3\alpha$ for
our polycrystalline samples. The specific heat was studied in a
Quantum Design PPMS calorimeter by means of a relaxation technique.
In the electrical transport experiments the samples were
investigated by four-probe $\rho$ measurements using an alternating
DC-current. The ESR measurements at a frequency of $\nu=9.6$\,GHz
were carried out in a standard Bruker EMX system. The HF-ESR
experiments were performed with a home-made spectrometer
\cite{golze_alfonsov_ni4} at frequencies $\nu=83-348$\,GHz and
magnetic fields $B=0-15$\,T. All ESR measurements were made in a
temperature range of $5-300$\,K.

\subsection{Sample preparation}

The polycrystalline samples  \GdOFFeAs\ ($x = 0, 0.15, 0.17,$
nominal content) were prepared by two different routes. Route~1,
which is similar to that described in
Ref.~[\onlinecite{Kondrat2009}], starts with FeAs, Gd, Gd$_2$O$_3$
and GdF$_3$ in a stoichiometric ratio. All materials were
homogenized by grinding in a mortar. Route~2 uses GdAs, Fe,
Fe$_2$O$_3$ and FeF$_3$ as starting materials in a stoichiometric
ratio. Here, the starting materials were homogenized by grinding in
a ball mill. In either case, the resulting powders were pressed into
pellets under Ar atmosphere, and subsequently annealed in an
evacuated quartz tube either in a two step synthesis at 940$^\circ$C
for 12~h and at 1150$^\circ$C for 48~h (60~h) or in a one step
synthesis at 940$^\circ$C for 168~h. In order to confirm the single
phase character of the polycrystals, powder x-ray diffraction was
performed on a Rigaku diffractometer (Cu $K_{\alpha}$-radiation,
graphite monochromator). The samples were either phase pure or
contained insignificantly small amounts of GdAs, GdOF, and
Fe$_3$O$_4$. The microstructure and the composition were examined by
scanning electron microscopy (SEM, XL30 Philipps, IN400) equipped
with an electron microprobe analyzer for semi-quantitative elemental
analysis using the wave length dispersive x-ray (WDX) mode. The
analysis showed that the sample with $x = 0.15$ (nominal content) in
fact contains $\sim 0.07\pm0.02$ of F and $x = 0.17$ contains $\sim
0.14\pm0.02$ of F. Further on, we will use the doping levels
obtained by WDX in order to label the samples.

\begin{figure}
\includegraphics[width=18pc, angle=0]{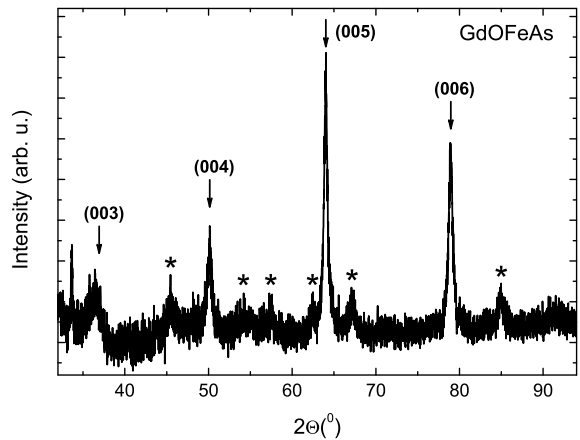}
\caption{Powder x-ray diffraction data of the $c$-axis aligned
GdOFeAs sample.} \label{fig:x-ray_GdOFeAs}
\end{figure}

For the $c$-axis alignment of the parent GdOFeAs sample, which was
synthesized under high pressure\cite{Zhigadlo2008}, the powder was
mixed with epoxy resin and hardened while rotating in a magnetic
field of $1.5$\,T. The x-ray diffraction data of the aligned powder
samples were collected at room temperature using a PANalytical
X'Pert PRO system (Philips) with Co $K_{\alpha}$-radiation
(\figref{fig:x-ray_GdOFeAs}). The presence of highly intense [00l]
reflections (\figref{fig:x-ray_GdOFeAs}, arrows) which dominate the
pattern points to a sufficiently good quality of the alignment.
Reflections with Miller indices different from [00l]
(\figref{fig:x-ray_GdOFeAs}, asterisks) are visible in the
background, too, but their intensity is strongly suppressed compared
to the powder pattern.

\section{Thermodynamic and transport measurements}
\label{s:tdandtr}

\begin{figure}
\includegraphics[width=20pc, angle=0]{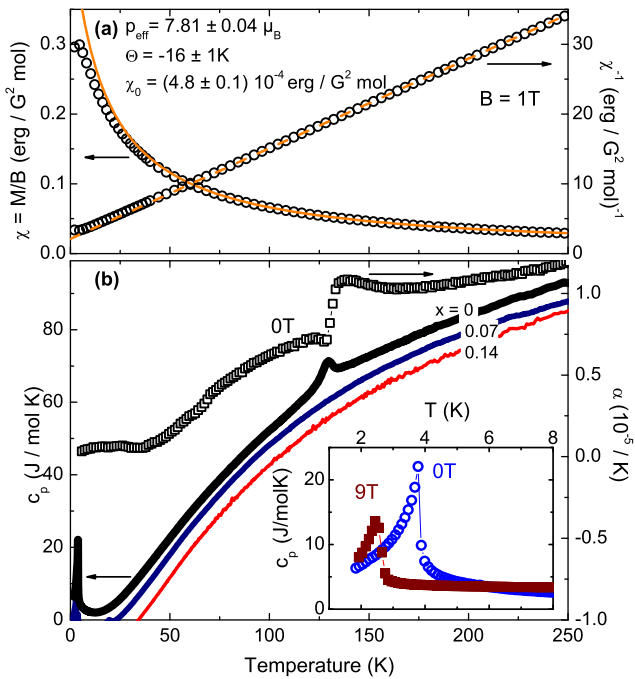}
\caption{(Color online) \textbf{(a)} Static susceptibility
$\chi=M/B$ (left axis), $\chi^{-1}$ (right axis); \textbf{(b)}
specific heat \cp\ (left axis) and thermal expansion coefficient
$\alpha$ (right axis) of \gdo\ vs. temperature; specific heat data
for $x=0.07$ and $x=0.14$ samples is artificially shifted down for
clarity; inset in \textbf{(b)} panel presents specific heat anomaly
associated with long-range antiferromagnetic ordering of the
Gd-moments at $B=0$ and $B=9$T. } \label{fig_x0_TD}
\end{figure}

In \figref{fig_x0_TD}(a), the temperature dependence of the static
susceptibility $\chi=M/B$ of \gdo\ is presented. As $\chi$($T$) is
dominated by the response of the Gd moments, the results are very
similar for the F-doped samples, which are not shown. In general,
the data obey the Curie-Weiss law which is expected due to the
presence of paramagnetic Gd$^{3+}$ ions. Note that the response of
the FeAs-layers which is e.g. visible in \lao\ is about 3 orders of
magnitude smaller and hence masked by the magnetism of the rare
earth ions.~\cite{Klingeler2010} The linear temperature dependence
of the inverse susceptibility demonstrates the Curie-Weiss-like
behavior. Analyzing the data in terms of the Curie-Weiss-law yields
the antiferromagnetic Weiss temperature $\Theta = -16\pm1$\,K and
the effective magnetic moments  $p_{\rm eff}= 7.81\pm0.04$~\mb\
which is close to the magnetic moment of a free Gd$^{3+}$ ion
($p_{\rm eff}= 7.94$~\mb). At a low temperature of about $\sim5$\,K
there is a kink of the magnetization due to the AFM ordering of the
Gd moments.

While the structural and Fe magnetic phase transitions are not
visible in the magnetization data, there are pronounced anomalies in
the specific heat \cp\ and the thermal expansion coefficient
$\alpha$ (\figref{fig_x0_TD}(b)) in the case of the parent GdOFeAs
sample. There is one broad feature visible in the specific heat
data. In contrast, the thermal expansion coefficient exhibits two
huge anomalies with opposite sign which can be attributed to the
structural and SDW transitions of the compound at $\Tst=136\pm 5$\,K
and $\Tsdw=128\pm 2$\,K. In addition, the specific heat data reveal
a sharp anomaly at \tngd\ = 3.8\,K which is associated with the
onset of long range antiferromagnetic order of the Gd moments, in
accord with the magnetization data. Note that the anomaly is not
present in our thermal expansion data due to the restricted
temperature range $T\geq 6$\,K. Upon application of external
magnetic fields, Gd order is strongly suppressed as shown in
\figref{fig_x0_TD} (inset in the lower panel). \tngd\ is shifted to
2.5\,K in an external magnetic field of $B=9$\,T. While anomalies
associated with Gd-ordering are still observed in the specific heat
data of the F-doped samples with $x=0.07$ and $x=0.14$, there are no
visible anomalies at higher temperatures (\figref{fig_x0_TD}(b)).
This evidences the absence of long-range SDW order in the doped,
superconducting samples.

\begin{figure}
\includegraphics[width=20pc, angle=0]{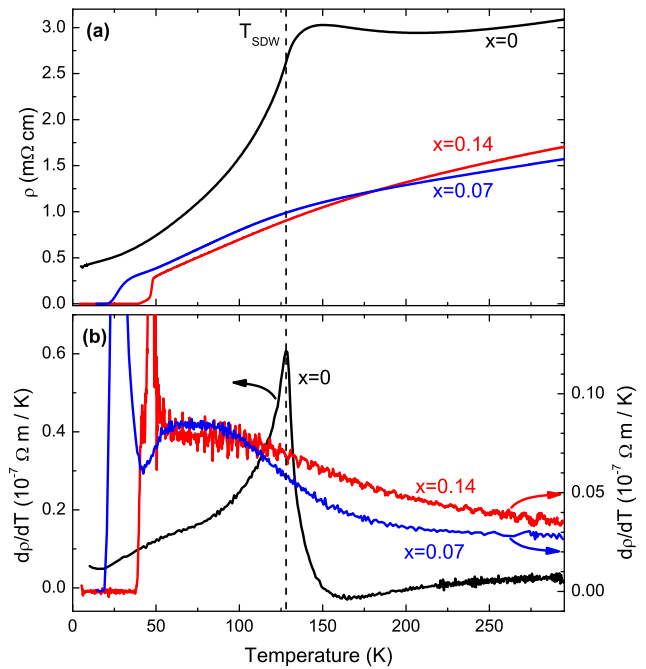}
\caption{(Color online) \textbf{(a)} Electrical resistivity $\rho$
of \GdOFFeAs\ samples, $x$ = 0, 0.07, 0.14; \textbf{(b)} electrical
resistivity derivative d$\rho{/}$dT for $x$ = 0 (left axis), $x$ =
0.07, 0.14 (right axis).} \label{fig:GdOFFeAs_resist}
\end{figure}

\figref{fig:GdOFFeAs_resist}(a) shows the temperature dependence of
the electrical resistivity $\rho$ of the \GdOFFeAs\ samples for all
three doping levels: $x=0$, 0.07 and 0.14. To get a better insight
into the data we present the temperature derivatives $d\rho/dT$ in
the bottom panel (\figref{fig:GdOFFeAs_resist}(b)). The resistivity
of the undoped material exhibits features closely connected to the
structural and magnetic phase transitions: a maximum close to
T$_{ST}$ and an inflection point at T$_{SDW}=128$~K, which are
characteristic for all the 1111 parent
compounds\cite{Hess2009,Klauss2008,Kondrat2009}. With doping, the
electrical resistivity drastically changes its behavior,
superconductivity emerges at low temperature and the intermediate
temperature maximum disappears. The SC temperatures \Tc\ for
$x=0.07$ and $x=0.14$ samples amount to $20$\,K and $45$\,K,
respectively. No pronounced features of the SDW phase are present in
these compounds in the whole investigated temperature range.
However, the normal state behavior of $\rho(T)$ for $x=0.07$ is very
unusual. At high temperatures the resistivity is linear down to
approx $200$~K, then it develops a curvature and drops below the
linear approximation of the high temperature part. With decreasing
temperature further, $\rho(T)$ becomes linear again and develops a
slight opposite curvature at $T\lesssim50$~K, prior to the onset of
superconductivity. Upon increasing the F doping level in the
samples, namely to $x=0.14$, this anomaly becomes weaker. A similar
drop of $\rho(T)$ at $T\lesssim200$~K as found here has previously
been observed for other 1111-type pnictide superconductors
\cite{Hess2009,Kondrat2009,Zhigadlo2010} as well as for
Ba$_{1-x}$K$_x$Fe$_2$As$_2$ \cite{Rotter2008a, Koitzsch2009}. The
qualitative resemblance to the sharp drop at $T_{SDW}$ which is
observed in the respective parent compounds suggests that the
resistivity drop in the superconducting samples is indicative of
remnants of the SDW phase. In fact, a recent study of the Nernst
effect on \LaOFFeAs\ provides strong evidence that precursors of the
SDW phase develop in the vicinity of the resistivity anomaly despite
the absence of static magnetism \cite{Kondrat2010}.

\section{Electron Spin Resonance, results and discussion }

\subsection{GdOFeAs, 9.6 GHz X-band measurements}

\begin{figure}
\includegraphics[width=18pc, angle=0]{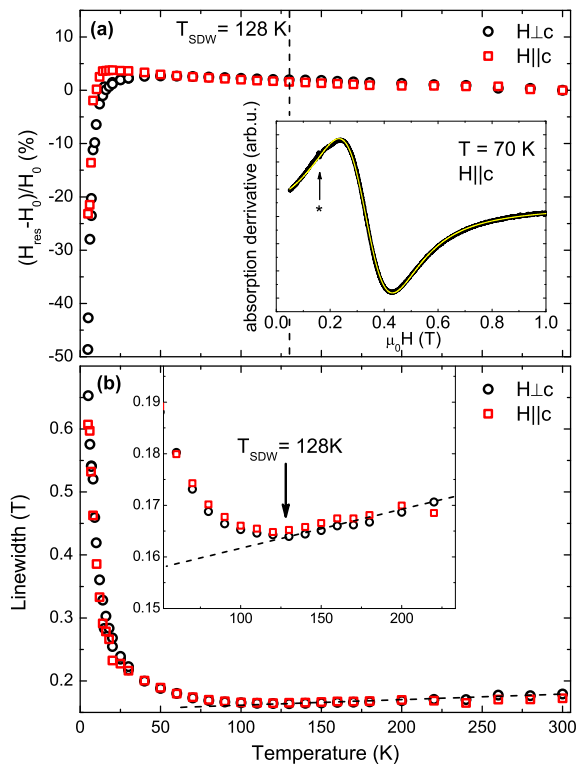}
\caption{(Color online) Results of the X-Band ESR ($\nu = 9.6$\,GHz)
measurements performed on $c$-axis aligned powder GdOFeAs sample for
two sample orientations in the magnetic field, open circles - field
in $ab$ plane, open squares - field parallel to the $c$-axis;
\textbf{(a)} temperature dependence of the resonance field on a
reduced field scale $(H-H_0)/H_0$. Here
$\mu_{0}H_{0}=0.323/0.328$\,T is resonance field at the highest
temperature for $H\perp c$/$H\parallel c$; the inset shows the
spectrum at $T=70$\,K in $H\parallel c$ configuration, the arrow
points at a small Fe ESR signal which presence indicates a small
amount of impurities; \textbf{(b)} temperature dependence of the
linewidth; inset shows the change of the T-dependence at the SDW
transition} \label{fig:GdOFeAs_xband}
\end{figure}

The ESR measurements performed at a frequency of $9.6$\,GHz on the
$c$-axis aligned GdOFeAs sample in the whole temperature range of
study and for both sample orientations (\Hext\ $||$ or $\perp c$)
reveal one broad line with the $g$-factor $\sim2$
(\figref{fig:GdOFeAs_xband}(a), inset). Such ESR response is typical
for the systems where \Gd3+\ ions occupy regular positions in the
crystal lattice with short distances between neighboring
ions\cite{barnes_esrinmetals}. The \Gd3+\ is an S-state ion with a
half-filled $4f$ shell which yields an isotropic $g$-factor equal to
2 and a spin value of 7/2. The rather big spin value leads to strong
magnetic dipole-dipole interactions which together with the
unresolved fine structure broaden the ESR line. This broadening
mechanism should lead to a gaussian line shape which is however not
observed in the spectra. Instead, the lorentzian\footnote{Note here
that a dysonian ESR line shape, which is typical for metallic
samples in a cavity, is not observed here due to the fine grinding
of the sample and further mixing with epoxy. This procedure reduces
the microwave dispersion which is the origin of the dysonian line
shape.} function had to be used to fit the spectra (thin line in the
inset in \figref{fig:GdOFeAs_xband}(a)) in order to obtain accurate
values of the resonance field ($H_{res}$) and the linewidth. The
lorentzian shape suggests that homogeneous narrowing of the line
does take place, which can be caused by isotropic exchange
interaction between Gd spins\cite{AbragamBleaney,
barnes_esrinmetals}.  The temperature dependencies of $H_{res}$ and
the linewidth are shown on \figref{fig:GdOFeAs_xband}(a),(b). With
lowering the temperature no drastic changes are seen in $H_{res}$
down to $\sim10$\,K where there is a strong shift of the line due to
the ordering of Gd moments (\figref{fig:GdOFeAs_xband}(a)). The
linewidth, in contrast to the resonance field, shows clear change in
the behavior at $\text{\Tsdw}=128$\,K for both sample orientations
(\figref{fig:GdOFeAs_xband}(b)). At temperatures above \Tsdw\ there
is a gradual decrease of the width of the ESR line upon cooling.
This can be attributed to a Korringa-like behavior, with the
linewidth having a linear in $T$ contribution due to the relaxation
of the Gd spins through interaction with the conduction electrons,
similar to EuFe$_2$As$_2$\cite{PhysRevB.81.024406}. The slope value
amounts to $\sim0.9\cdot10^{-4}$\,T/K which is one order of
magnitude smaller than that in EuFe$_2$As$_2$.  A strong broadening
of the line below \Tsdw\ can be attributed to the formation of the
SDW state in the FeAs layers. Similar effects in the \Gd3+\ ESR
linewidth were observed before in the case of Gd$_2$BaCuO$_5$
samples where exchange coupling of Gd- and Cu-moments enabled to
probe by means of \Gd3+\ ESR the magnetic ordering of the Cu
layers\cite{goya_R2BaMO5_ESR} (see the discussion below).

\subsection{GdOFeAs, high-frequency/field measurements}

\begin{figure}
\includegraphics[width=20pc, angle=0]{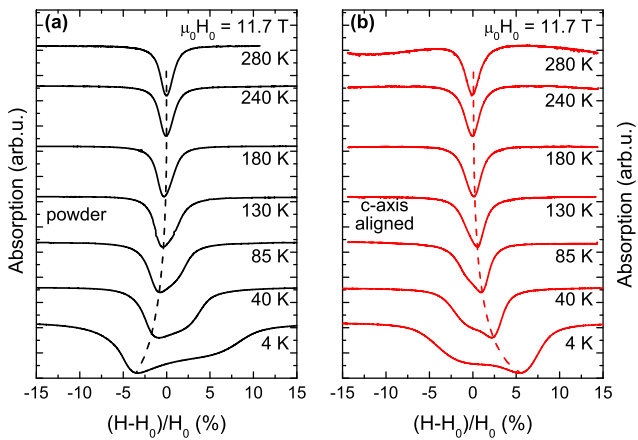}
\caption{(Color online) Temperature evolution of the
high-frequency/field ESR spectra of GdOFeAs powder and $c$-axis
aligned powder samples at a frequency of $\nu = 328$\,GHz, shown on
a reduced field scale $(H-H_0)/H_0$. Here $\mu_0 H_0 =11.7$\,T is
the resonance field of the signal at high temperature; \textbf{(a)}
non-oriented powder; \textbf{(b)} $c$-axis oriented powder.}
\label{fig:tdepsp_GdOFeAs}
\end{figure}

In the measurements performed at 9.6\,GHz, the linewidth of the
\Gd3+\ ESR signal is comparable to its resonance field. This leads
to complications in the spectra analysis and to a lack of
resolution. In order to improve the spectral resolution we performed
high-frequency/field measurements on GdOFeAs samples. The
high-temperature ESR spectra of the non-oriented GdOFeAs powder
sample measured at a frequency of 328\,GHz
(\figref{fig:tdepsp_GdOFeAs}(a)) consist of a single broad
lorentzian-shaped line with a $g$-factor of $\sim2$ and a linewidth
of $\sim0.2$\,T, similar to the low-frequency measurements. However,
the very small Korringa contribution detected in the low frequency
measurements is not visible in the HF-ESR spectra
(\figref{fig:LW_GdOFFeAs}a). The low-temperature HF-ESR spectra
exhibit an inhomogeneously broadened shape which is in contrast to
X-band data. As a measure of this broadening the full width at the
half maximum (FWHM) of the signal $\Delta{H}$ has been taken (see
\figref{fig:LW_GdOFFeAs}(a)). As can be seen, with decreasing the
temperature there is only a weak broadening of the signal down to a
characteristic temperature $\Tif\sim150$\,K ($if$ denotes an
internal field at the Gd ion, see the discussion below) where
substantial inhomogeneous broadening begins to develop continuously
down to the lowest measured temperature of $4$\,K.
\begin{figure}
\includegraphics[width=18pc, angle=0]{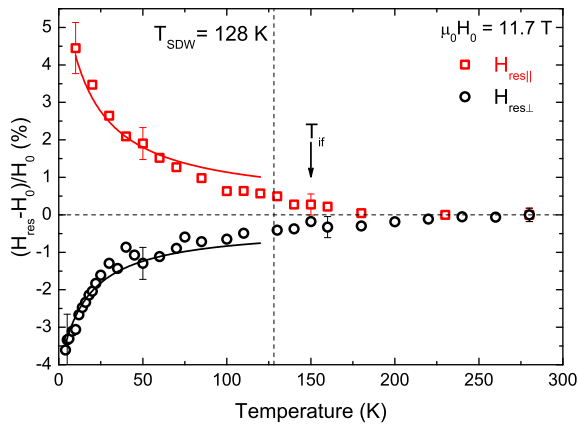}
\caption{(Color online) The shift of the minimum of absorption of
the spectra with temperature, measured at a frequency of $\nu =
328$\,GHz,  on a reduced field scale $(H_{res}-H_0)/H_0$. Here
$\mu_0 H_0 =11.7$\,T is the resonance field of the signal at high
temperature; open circles - powder sample, open squares - $c$-axis
aligned powder sample, solid lines - fits of the resonance field
using Eq.~\ref{Hgd4}.} \label{fig:tdeprf_GdOFeAs}
\end{figure}
Concomitantly with the inhomogeneous broadening there is a
noticeable shift of the minimum of the absorption to lower fields,
as shown on \figref{fig:tdeprf_GdOFeAs} (open circles) on a reduced
field scale $(H-H_0)/H_0$. Here $\mu_0 H_0 =11.7$\,T is the
resonance field of the signal at high temperature. The spectral
shape at low temperature (\figref{fig:tdepsp_GdOFeAs}(a)) appears to
be very similar to the shape of the ESR signal from a powder sample
with an anisotropic $g$-factor. However, since the Gd$^{3+}$ ion is
a pure S-state ion, it should have an isotropic $g$-factor very
close to 2. Hence one can conjecture that the shape of the ESR
signal from the GdOFeAs sample is caused by the anisotropy of the
internal field at the Gd site arising from the AFM-ordered Fe
moments. In such an anisotropic powder situation most of the
spectral weight is coming from the grains whose $c$-axes are
oriented perpendicular to the direction of the external field
$H_{\rm ext}$. One should note here that in the case of in-plane
anisotropy there will be an additional averaging effect due to the
distribution of resonance fields of grains whose ($ab$)-planes are
parallel to $H_{\rm ext}$. Therefore we assume here that the
low-field minimum of the absorption corresponds to the mean value of
the resonance field of the \Gd3+\ ESR response (\Hrespp) in the case
of the external field applied perpendicular to the $c$-axis.
Correspondingly, the high-field shoulder of the spectra arises from
grains whose $c$-axes make small angles with respect to $H_{\rm
ext}$.

In order to probe the \Gd3+\ response for the geometry $H_{\rm
ext}\parallel c$ we have performed ESR measurements on the $c$-axis
oriented GdOFeAs powder sample. Though, according to the x-ray
diffraction analysis, the alignment of the powder particles was not
perfect, a substantial $c$-axis texturing of the sample has been
achieved (\figref{fig:x-ray_GdOFeAs}). Similarly to the non-oriented
powder sample, the $c$-axis oriented sample at temperatures above
$\sim 150$\,K exhibits only a small broadening of the ESR spectrum
with decreasing temperature (\figref{fig:tdepsp_GdOFeAs}(b)). Below
this temperature the signal experiences strong inhomogeneous
broadening where most of the spectral weight is shifted to higher
fields, which is opposite to the finding in the non-oriented powder
sample. In the oriented sample most of the spectral weight and,
consequently, the minimum of the absorption should correspond to the
resonance field of the \Gd3+\ ESR response (\Hrespl) in the geometry
\Hext\,$\parallel c$ whereas a non-ideal powder alignment yields the
low-field shoulder of the ESR signal.

From our measurements on non-oriented and $c$-axis oriented powder
samples one can, therefore, extract the temperature dependencies of
$H_{res}$ in two configurations, i.e., for fields aligned along the
$c$-axis (\Hrespl) and in the ($ab$)-plane (\Hrespp), as summarized
in \figref{fig:tdeprf_GdOFeAs}. As can be seen, the changes of both
resonance fields \Hrespp\ and \Hrespl\ start upon cooling at
$\Tif\lesssim150$\,K and the shifts have opposite directions.

\begin{figure}
\includegraphics[width=20pc, angle=0]{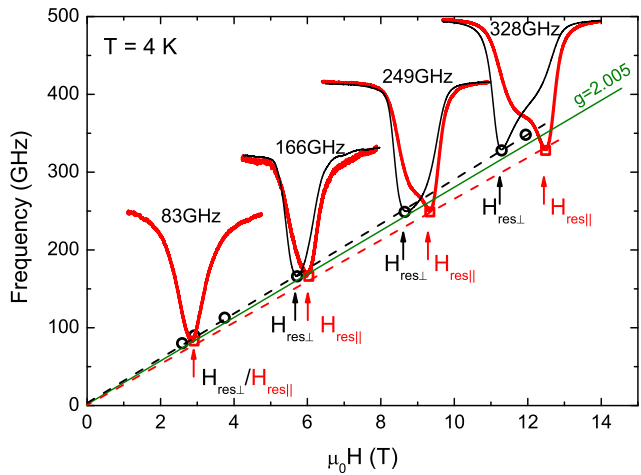}
\caption{(Color online) Frequency dependence of the ESR signal from
GdOFeAs non-oriented powder and $c$-axis oriented powder samples
measured at $T = 4$\,K; data points correspond to the position of
the minimum of absorption of the spectra at different frequencies,
open circles - non-oriented powder sample, open squares - $c$-axis
oriented powder sample, dashed lines - guides for the eyes, solid
line - the frequency dependence of the position of the high-T Gd ESR
line measured at $T = 280$\,K; the lineshape of the spectra is shown
as well, thin solid line - non-oriented powder, thick solid line -
$c$-axis oriented powder.} \label{fig:fdep_GdOFeAs}
\end{figure}

The qualitative difference in the high-field/frequency and
low-field/frequency measurements leads to the conclusion that the
shift of the resonance field and the inhomogeneous broadening of the
spectra measured at a frequency of 328\,GHz is a field-induced
effect. To investigate it, we have measured the frequency $\nu$
versus magnetic field $H_{\rm res}$ dependence of the GdOFeAs powder
and the $c$-axis aligned samples, respectively, both at $T=280$\,K
and $T=4$\,K (\figref{fig:fdep_GdOFeAs}). At $T=280$\,K the spectrum
at all studied frequencies and fields consists of a single
lorentzian line with the same linewidth value. A linear $\nu
(H_{res})$ dependence has been revealed yielding a $g$-factor $g =
h\nu/(\mu_BH_{\rm res})$ equal to 2.005 (straight solid line on
\figref{fig:fdep_GdOFeAs}). The spectra at $T=4$\,K for the
non-oriented powder (thin line) and $c$-axis oriented powder (thick
line) samples at different frequencies together with the frequency
dependence of the resonance fields \Hrespp\ and \Hrespl\ are shown
on \figref{fig:fdep_GdOFeAs} as well. This measurement reveals that
the difference between \Hrespp\ and \Hrespl\ increases linearly with
increasing the frequency and the field strength.

\subsection{GdOFeAs, discussion}

Since the inhomogeneous broadening and the shift of the Gd ESR
signal set in close to the SDW ordering temperature, one can
associate them with the interaction of the Fe ordered moments with
the Gd spins. The shifts of \Hrespp\ and \Hrespl\ from a common
high-temperature paramagnetic value $H_{\rm 0}$ can hence be related
to the occurrence of an internal magnetic field (\Hintpp\ and
\Hintpl) at the Gd site due to the formation of the static SDW in
the FeAs layer. Based on the dependence shown on
\figref{fig:fdep_GdOFeAs} we can conclude that the strength of these
internal fields depends on the strength of the applied magnetic
field.

\begin{figure}
\includegraphics[width=20pc, angle=0]{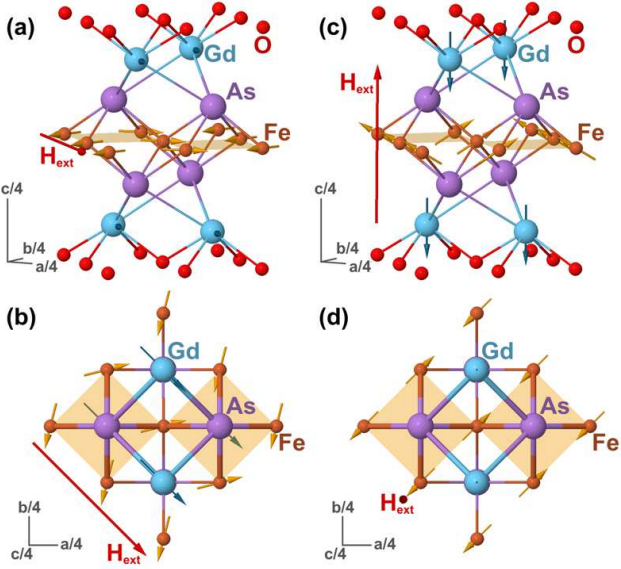}
\caption{(Color online) Canting of the Fe moments due to the applied
magnetic field \Hext; \textbf{(a),(b)} \Hext$\perp c$ (angle between
\Hext\ and Fe AFM ordered moments is $90^\circ$); \textbf{(c),(d)}
\Hext$\parallel c$; arrows on the Fe site depict magnetic moments,
whereas arrows on the Gd sites represent the induced internal
field.} \label{fig:GdOFeAs}
\end{figure}

When the FeAs planes are in the paramagnetic state, at temperatures
above $\Tsdw=128$\,K, the exchange/dipolar field at the Gd position
is negligible as the applied magnetic field cannot effectively
polarize small paramagnetic moments at elevated temperatures. At
temperatures below $\Tsdw=128$\,K the Fe moments order
\emph{statically} in the ($ab$)-plane. The AFM structure of the
magnetic order, which is similar for all $R$-based 1111-type
pnictides can be found in Ref.~\onlinecite{maeter_feas_musr_2009}.
Due to symmetry reasons, without an applied magnetic field the
internal field at the Gd site can be only of dipolar nature and in
this case according to an estimate it should not exceed
$\sim0.03$\,T. We suggest that application of an external magnetic
field induces tilting of the Fe spins and hence creates an
uncompensated magnetic moment ($m_{Fe}$) in the direction of the
external field\footnote{For simplicity which does not affect the
conclusions of the following qualitative discussion, we assume
hereafter that the uncompensated magnetic moments and thereby
created internal fields are collinear with the applied magnetic
field.}. This moment can interact with the \Gd3+\ spins which would
effectively lead to the occurrence of an additional internal field.
If \Hext\,$\perp c$, then the Fe spins tilt in the $ab$ plane as it
is schematically shown on \figref{fig:GdOFeAs}(a,b) for the
configuration where \Hext\ makes an angle of $90^\circ$ with the AFM
ordered Fe spins. The shift of the ESR line \Hrespp, measured at
328\,GHz, yields an estimate of the internal field (\Hintpp) of
about $\approx0.4$\,T parallel both to \Hext\ and to the
uncompensated moment $m_{Fe}$ in the FeAs plane. If
\Hext\,$\parallel c$, then the Fe moments tilt out of the plane
(\figref{fig:GdOFeAs}(c,d)). In this case the shift of the ESR line,
measured at 328\,GHz, yields an estimate of the internal field
(\Hintpl) of about $\approx0.65$\,T antiparallel both to the \Hext\
and to the uncompensated moment $m_{Fe}$. One should note here that
an estimate of the dipolar field produced by the FeAs layers at the
Gd site yields a value which does not exceed $\approx0.05\,$T even
for full out-of-plane canting. This field is one order of magnitude
smaller than the experimentally observed value which clearly implies
the presence of an appreciable exchange interaction between Gd and
canted Fe moments. The dependence of the sign of the internal field
on the direction of the applied magnetic field suggests that the
sign of the exchange interaction with Gd spins is different for
different directions of the uncompensated Fe moments, i.e.,
ferromagnetic for the in-plane and antiferromagnetic for the
out-of-plane directions. This surprisingly strong anisotropy of the
exchange might be related with the multiband electronic structure of
iron pnictides which might give rise to different pathways for
interactions between the Gd 4$f$ orbitals and the in-plane $xy$ and
out-of-plane $xz$ and $yz$ Fe 3$d$ bands. Note that, in zero
magnetic field and hence without Fe spin canting ($m_{Fe}=0$), the
exchange interaction between Fe and Gd moments of an arbitrary sign
is geometrically frustrated (see \figref{fig:GdOFeAs}). The
application of a field which tilts the Fe moments thus removes this
frustration.

Considering the exchange interaction, one might suppose that the
temperature dependence of the internal field at the Gd site should
follow the behavior of the SDW order parameter in 1111
compounds\cite{maeter_feas_musr_2009, Luetkens2009, Cruz2008_feas},
which increases fast within $\sim30$\,K starting at \Tsdw\ and then
stays almost constant with further decreasing the temperature. Here,
the internal field acting on the \Gd3+\ moments arises at a
temperature $\Tif\approx150$\,K which is $\sim20$\,K higher than
$\Tsdw=128$\,K and keeps increasing upon cooling till the lowest
measured temperature (see \figref{fig:tdeprf_GdOFeAs}). When
approaching the temperature of the SDW transition from high
temperatures, the appearance of the internal field well above \Tsdw\
can be explained by growing quasi-static correlations between the Fe
moments seen in the time window of the high-frequency ESR. The
development of the internal field below \Tsdw\ is found to be
similar to that of some other systems where paramagnetic ions are
coupled to magnetically-ordered moments of another type
\cite{PhysRevB.50.16708, PhysRevB.41.1934}. To explain this
evolution of the internal field we use a simple model based on a
mean-field approximation\cite{Kittel, barnes_esrinmetals} (for
details see Appendix). According to this model the internal field
depends on the applied magnetic field and on temperature as
following (see Eq.~\ref{Hgd5}):
\begin{eqnarray}\label{Hgd4simp}
\nonumber H_{int} \sim \pm\frac{A(J, \alpha,
C_{Gd})\text{\Hext}}{(T-\Theta)}
\end{eqnarray}
Here $A(J, \alpha, C_{Gd})$ is a parameter determined by the Gd-Fe
exchange coupling energy $J$, by the susceptibility of the ordered
Fe moments to the applied field $\alpha$ and by the Gd Curie
constant $C_{Gd}$. As can be seen, this dependence qualitatively
obeys a Curie-Weiss law at a given applied field which agrees well
with the measured data. The model enables to fit the experimental
data points reasonably well (see Appendix and
\figref{fig:tdeprf_GdOFeAs}). Firstly, the fit yields an estimate of
the energy of the exchange coupling between Gd and Fe spins which is
in the range of $|J|\approx 15-20$\,K. Secondly, it shows that the
magnitude of the uncompensated Fe moment depends on the Gd
magnetization which suggests that the Gd subsystem additionally
tilts or polarizes the SDW.

To summarize this part, the ESR results on GdOFeAs samples show that
the Gd subsystem is exchange-coupled to the magnetic FeAs planes. On
approaching the AFM SDW transition from above, the growing
correlations between the Fe moments yield a shift of the Gd ESR
line. At lower temperature, depending on the angle between \Hext\
and the $c$-axis of the sample, the signal shifts to higher or to
lower fields due to the uncompensated exchange field which is
transferred to the Gd site from the Fe moments canted in an external
magnetic field. Since the full width at the half maximum $\Delta{H}$
is proportional to the difference between resonance fields
$\text{\Hrespp}-\text{\Hrespl}$, then the width of the ESR signal
$\Delta{H}$ of the non-oriented powder sample can be taken as a
measure of this exchange field (\figref{fig:LW_GdOFFeAs}a).

\subsection{\GdOFFeAs\ ($x=0.07$, $0.14$)}

\begin{figure}
\includegraphics[width=18pc, angle=0]{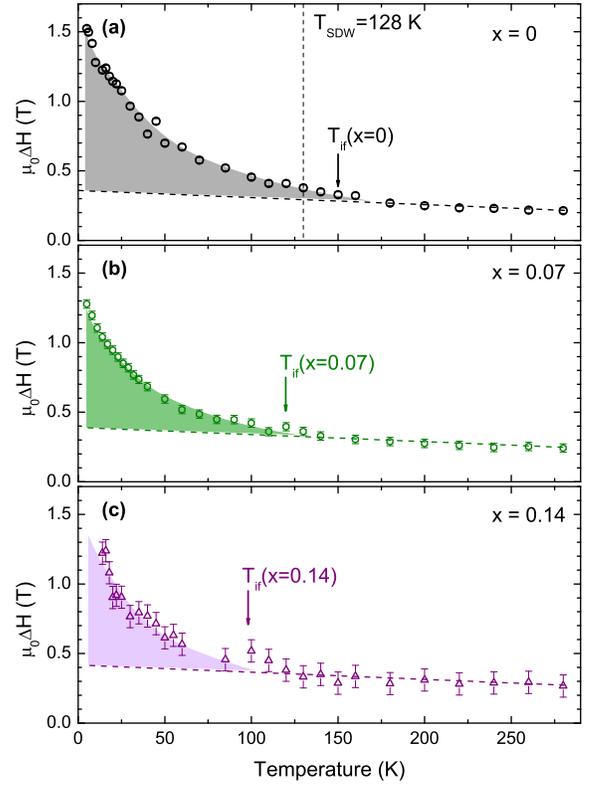}
\caption{(Color online) The full width at the half maximum (FWHM)
$\Delta{H}$ of the ESR lines, measured at a frequency of $\nu =
348$\,GHz ($x = 0.07$) and $\nu = 328$\,GHz ($x = 0, 0.14$), as a
function of temperature for the non-oriented \GdOFFeAs\ samples;
\textbf{(a)} $x = 0$; \textbf{(b)} $x = 0.07$; \textbf{(c)} $x =
0.14$.} \label{fig:LW_GdOFFeAs}
\end{figure}
\begin{figure}
\includegraphics[width=20pc, angle=0]{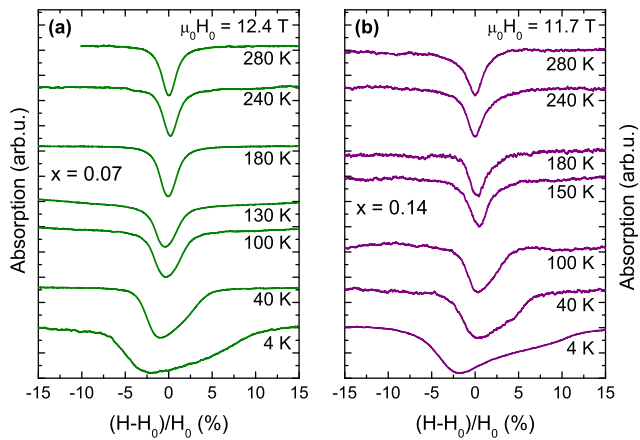}
\caption{(Color online) Temperature evolution of the
high-frequency/field ESR spectra of \GdOFFeAs\ powder samples
measured at a frequency of $\nu = 348$\,GHz ($x = 0.07$) and $\nu =
328$\,GHz ($x = 0.14$), shown on a reduced field scale
$(H-H_0)/H_0$. Here $H_0$ is the resonance field of the signal at
high temperature; \textbf{(a)} $x = 0.07$, $\mu_0 H_0 =12.4$\,T;
\textbf{(b)}  $x = 0.14$, $\mu_0 H_0 =11.7$\,T.}
\label{fig:tdepsp_GdOFFeAs}
\end{figure}
\begin{figure}
\includegraphics[width=18pc, angle=0]{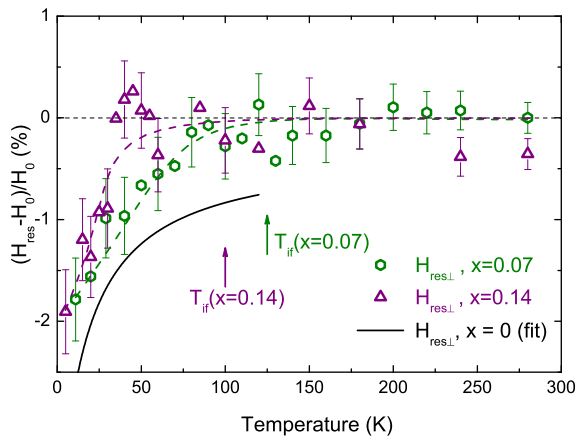}
\caption{(Color online) The shift of the minimum of absorption of
the spectra with temperature, measured at a frequency $\nu =
348$\,GHz($x = 0.07$)/328\,GHz($x = 0.14$), on a reduced field scale
$(H_{res}-H_0)/H_0$. Here $\mu_0 H_0 =12.4$\,T/11.7\,T is the
resonance field of the signal at high temperature, open hexagons -
$x = 0.07$, open triangles - $x = 0.14$, solid line - fit of the
resonance field in the parent GdOFeAs sample.}
\label{fig:tdeprf_GdOFFeAs}
\end{figure}

\begin{figure}
\includegraphics[width=20pc, angle=0]{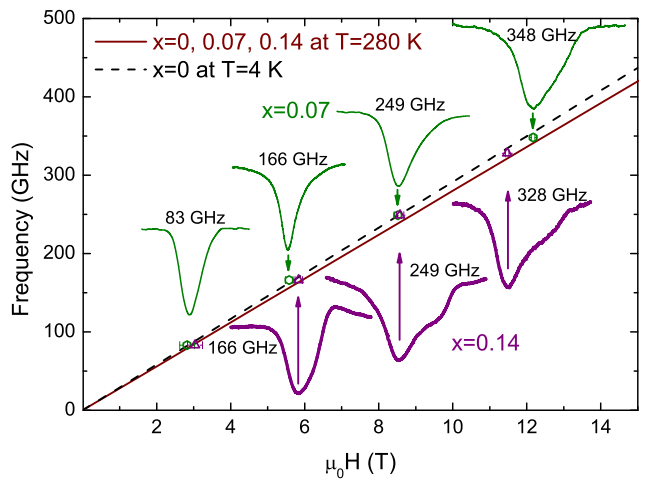}
\caption{(Color online) Frequency dependence of the ESR signal from
\GdOFFeAs\ powder samples measured at a temperature of $T = 4$\,K,
in SC state; data points correspond to the position of the minimum
of absorption of the spectra at different frequencies, open hexagons
- $x = 0.07$, open triangles - $x = 0.14$, dashed line - guide for
the eyes, solid line - the frequency dependence of the position of
the high-T Gd ESR line measured at $T = 280$\,K for both F-doped
samples; the lineshape of the spectra is shown as well, thin solid
line - $x = 0.07$, thick solid line - $x = 0.14$.}
\label{fig:fdep_GdOFFeAs}
\end{figure}

The influence of the fluorine doping on the Gd ESR has been studied
on two powder samples of \GdOFFeAs\ with 7\% and 14\% of fluorine.
On \figref{fig:tdepsp_GdOFFeAs}(a),(b) the evolution of the
respective \Gd3+\ ESR spectra is shown on a reduced field scale.
Similarly to the undoped sample, at high temperature the ESR
spectrum for both doped samples consists of a single
lorentzian-shaped line with $g=2.005$. While lowering the
temperature, the line remains almost unchanged until a
characteristic temperature \Tif\ is reached. This temperature
corresponds to the onset of an additional inhomogeneous contribution
to the width of the ESR signal $\Delta{H}$. This inhomogeneous
contribution is shown by the shaded area on
\figref{fig:LW_GdOFFeAs}(b,c). The temperature \Tif\ clearly depends
on the fluorine doping level. In the case of the 7\% F-doped sample
a noticeable broadening of the line starts at
\Tif$(x=0.07)\sim125$\,K, whereas for the 14\% doped sample it
starts at a lower temperature \Tif$(x=0.14)\sim100$\,K. For both
doped samples there is a shift of the minimum of the absorption
(\Hrespp) to lower magnetic fields below \Tif\
(\figref{fig:tdeprf_GdOFFeAs}). Qualitatively this shift is similar
to that of the undoped sample (solid line on
\figref{fig:tdeprf_GdOFFeAs}), but it is less pronounced. In
addition, the inhomogeneous broadening and the shift of \Hrespp\ to
lower magnetic fields exhibit a magnetic field dependence similar to
that of the undoped GdOFeAs sample (\figref{fig:fdep_GdOFFeAs}).

\subsection{Superconducting \GdOFFeAs, discussion}

The remarkable similarities of the Gd ESR behavior between the
fluorine-doped samples and the undoped one strongly suggest that,
even in the superconducting samples where the phase transition to
the AFM SDW state is not observed in the thermodynamics and
transport properties, quasi static (on the time scale of the ESR
measurement) magnetic correlations in the FeAs planes are present
below the characteristic temperature \Tif. Such correlations may
explain the peculiar features in the resistivity data shown in
Sec.~\ref{s:tdandtr}.

The unified phase diagram for the iron pnictides, especially for the
1111 materials, is not fully established so far since the issue of
coexistence of superconductivity and magnetism  remains
controversial. The nonmagnetic rare earth based \LaOFFeAs\ material
exhibits no evidence for the presence of a static magnetic order for
any superconducting composition, but rather reveals SDW-like spin
fluctuations seen in the transport\cite{Kondrat2010} and inelastic
neutron scattering experiments\cite{Shamoto2009, JPSJ.79.074715}. On
the other hand, the \emph{magnetic} rare earth based systems (Sm,
Nd, Ce) studied so far demonstrate coexistence of superconductivity
and static magnetism at least in the underdoped
region\cite{Drew2009, Carlo2009}. Our high-field ESR results show
that there is yet another 1111 system comprising a strongly magnetic
rare earth (Gd) subsystem where the coexistence of quasi-static
magnetism and superconductivity is still visible in large doping
range. Here, the increase of the fluorine content and
correspondingly the rise of \Tc\ leads to the suppression of
magnetic correlations indicating a possible interplay between these
two states. All this suggests that the coexistence and possible
interplay of the static or quasi-static magnetism and
superconductivity may be a generic property of 1111-type compounds.
In this regard, a remaining question yet to be answered is the
extent to which the $R$-Fe magnetic interaction influences the
magnetic correlations in the FeAs planes. Moreover, the issue of the
coexistence of magnetism and superconductivity is frequently
discussed in the literature for other pnictide families as well.
Extending ESR experiments to compounds of these families would be of
great interest.

\section{Conclusion}
Our HF-ESR study of polycrystalline samples of the \GdOFFeAs\
superconductor reveals a magnetic coupling between the Gd subsystem
and the FeAs layers. This coupling, most probably of the anisotropic
exchange type, is visible in the Gd ESR response in the undoped
GdOFeAs in the SDW state, in form of a field-induced inhomogeneous
broadening and shift of the ESR spectrum. This effect is caused by
the interaction of the Gd spins with the uncompensated Fe moments
due to the canting of the Fe moments in magnetic field. Furthermore,
the data suggest that the Gd moments additionally tilt the ordered
Fe moments. Surprisingly, the broadening and the shift of the
spectrum are present also in the doped superconducting samples where
there is no evidence of long range magnetic order. This points to
the presence of short range, static on the ESR time scale, magnetic
correlations. This may be relevant to the interplay of magnetism and
superconductivity in these materials, where on doping with fluorine
there is a simultaneous increase of the superconducting critical
temperature and suppression of the magnetic correlations. The
possible relevance of the exchange interaction between the magnetic
rare-earth subsystem and the FeAs planes to the properties of this
novel class of superconductors remains to be elucidated.

\begin{acknowledgments}

We thank S. M\"uller-Litvanyi, R. M\"uller, J. Werner, and S. Pichl
for assistance in the sample preparation. We thank U. Stockert and
J. E. Hamann-Borrero for assistance in the sample characterization.
The work at the IFW Dresden was supported by the Deutsche
Forschungsgemeinschaft through Grants No. BE1749/12 and BE1749/13,
the Research Unit FOR538 (Grant No. BU887/4) and the Priority
Programme SPP1458 (Grant No. GR3330/2). Work at the ETH was
supported by the Swiss National Science Foundation through the
National Center of Competence in Research MaNEP (Materials with
Novel Electronic Properties).

\end{acknowledgments}

\appendix*
\section{}

Here we provide a calculation of the internal field on the Gd ion
using a simple model based on the mean-field
approximation\cite{Kittel, barnes_esrinmetals}. To simplify the
calculations we assume that a magnetic field and magnetic moment
vectors are collinear. In this model, the internal field at the Gd
site $H_{int}$ is proportional to the magnitude of the
uncompensated Fe moment $m_{Fe}$ with a coefficient $\lambda$:
\begin{eqnarray}\label{Hgd1}
H_{int} = \pm\lambda  m_{Fe}
\end{eqnarray}
Hereafter the sign depends on the type of interaction, being "$+$"
for ferromagnetic and "$-$" for antiferromagnetic exchange.
Neglecting the weak dipolar contribution, the magnetization
normalized to the single ion $m_{Gd}$ of the Gd subsystem is
proportional to the sum of the applied magnetic field \Hext\ and
internal field $H_{int}$:
\begin{eqnarray}\label{Hgd2}
m_{Gd} = \chi_{Gd}(T)\, (\text{\Hext} \pm H_{int})\text{,}
\end{eqnarray}
where $\chi_{Gd}(T)=C_{Gd}/(T-\Theta)$ is the Gd magnetic
susceptibility, $C_{Gd}$ is the Gd Curie constant, $\Theta$ is the
Gd Curie temperature. Due to the exchange interaction, the
uncompensated magnetic moment $m_{Fe}$ is proportional not only to
the applied field but also to the internal field created by the Gd
moments:
\begin{eqnarray}\label{Hgd3}
m_{Fe} = \alpha (\text{\Hext} + \lambda m_{Gd})
\end{eqnarray}
Here $\alpha$ is the susceptibility of the ordered Fe moments to the
external magnetic field. Using Eq.~\ref{Hgd1}, Eq.~\ref{Hgd2} and
Eq.~\ref{Hgd3} one can obtain the equation for the internal field
$H_{int}$ at the Gd site:
\begin{eqnarray}\label{Hgd4}
H_{int} = \pm \frac{ \lambda^{2} \alpha C_{Gd} + \lambda
\alpha(T-\Theta) }{ (T-\Theta) - \lambda^{2} \alpha C_{Gd} }
\text{\Hext}
\end{eqnarray}

Eq.~\ref{Hgd4} enables to fit the measured temperature dependence of
the internal field (see \figref{fig:tdeprf_GdOFeAs}). The resonance
field of the Gd is determined by the applied field and by the
internal field $H_0 = \text{\Hext} \pm H_{int}$. At high
temperatures when there is no internal field at the Gd site at any
strength of the applied magnetic field ($H_{int}=0$) one can measure
the resonance field $H_0=11.7$\,T (for measurement frequency
$\nu=328$\,GHz). Assuming that the resonance field of the Gd ions
$H_0=11.7$\,T stays constant at all measured temperatures one
obtains the expression for the applied field $\text{\Hext}=H_0 \mp
H_{int}$ ($H_0 = 11.7$\,T). The fit for two measurement
configurations (\Hrespp\ and \Hrespl) is shown on
\figref{fig:tdeprf_GdOFeAs} by solid lines. The parameters $C_{Gd}$,
$\Theta$ and $\alpha$ can be taken from different experiments. The
Gd Curie constant $C_{Gd}$ and Curie temperature $\Theta$ are known
from the susceptibility data of GdOFeAs samples (see
Sec.~\ref{s:tdandtr}).
As it is shown in Ref.~\onlinecite{Klingeler2010} the bulk Fe
susceptibility of LaOFeAs samples is determined by the spin
susceptibility. Therefore the parameter $\alpha$ can be estimated
from this measurement yielding a value of $\sim10^{-4}$
$\frac{\text{erg}}{\text{G$^2$ mol}}$ \cite{Klingeler2010,
McGuire2008, Nomura2008, Kohama2008a}. The $\lambda$ value resulting
from the fit is equal to $\sim19.7$ $\frac{\text{G$^2$
mol}}{\text{erg}}$ for \Hrespp\ and $\sim25.3$ $\frac{\text{G$^2$
mol}}{\text{erg}}$ for \Hrespl. According to the mean field
theory\cite{Kittel}, these values yield an estimate of the exchange
interaction energy $J$ for two configurations amounting to $|J| \sim
15$\,K for \Hrespp\ and $|J| \sim 19$\,K for \Hrespl. In addition,
Eq.~\ref{Hgd1} enables to calculate the uncompensated moment
$m_{Fe}$. Its value grows with decreasing the temperature and
increasing the Gd susceptibility until it reaches $\sim0.03\mu_{B}$
at the lowest measured temperature.

Since $\alpha$ is very small compared to $C_{Gd}$ and $\lambda$,
Eq.~\ref{Hgd4} can be simplified to the form:
\begin{eqnarray}\label{Hgd5}
H_{int} \sim \pm \frac{ \lambda^{2} \alpha C_{Gd} }{ (T-\Theta)  }
\text{\Hext}
\end{eqnarray}

\bibliographystyle{apsrev_nourl_noissn}
\bibliography{alfonsov}

\end{document}